\documentclass[final,3p,times,twocolumn]{elsarticle}

\pdfoutput=1

\usepackage{lineno}

\usepackage{graphicx}
\usepackage{epstopdf}

\usepackage{amssymb}
\usepackage{amsmath}

\newcommand{\cd}{C$_6$D$_6$ }
\newcommand{\ngr}{(n,$\gamma$) }

\newcommand{\lacl}{LaCl$_{3}$ }

\journal{Nucl. Instr. and Meth. in Phys. Res. A}
\begin{document}

\begin{frontmatter}

\title{i-TED: A novel concept for high-sensitivity (n,$\gamma$) cross section measurements}

\author{C.~Domingo-Pardo}
\address{Instituto de F{\'\i}sica Corpuscular, CSIC-University of Valencia, Spain}


\begin{abstract}
A new method for measuring \ngr cross sections aiming at enhanced signal-to-background ratio is presented. This new approach is based on the combination of the pulse-height weighting technique with a total energy detection system that features $\gamma$-ray imaging capability (i-TED). The latter allows one to exploit Compton imaging techniques to discriminate between true capture $\gamma$-rays arising from the sample under study and background $\gamma$-rays coming from contaminant neutron (prompt or delayed) captures in the surrounding environment. A general proof-of-concept detection system for this application is presented in this article together with a description of the imaging method and a conceptual demonstration based on Monte Carlo simulations.
\end{abstract}

\begin{keyword}
radiative neutron capture \sep neutron time-of-flight \sep cross section \sep pulse-height weighting technique \sep Compton imaging



\end{keyword}

\end{frontmatter}

\section{Introduction}\label{sec:introduction}
The two most common methods for measuring \ngr cross sections as a function of
the neutron time-of-flight (neutron energy equivalent) employ either a high-efficiency total absorption
calorimeter (TAC)~\cite{Wisshak89,Ullmann05,Guerrero09} or a low efficiency setup of \cd
total energy detectors (TED)~\cite{Moxon63,Macklin67,Abbondanno04,Borella07}.
With either type of detection system, only the deposited energy (and multiplicity in the case of TAC) of the incoming
radiation can be measured, which implies that background rejection and hence, the optimization of the peak-to-back\-ground ratio, has to rely on appropriate selections or software-cuts of those quantities.

A new detection system is proposed in the present article, which allows one
to implement a further level of background rejection based on the spatial
origin (or incoming direction) of the measured $\gamma$-rays. To this aim, the
low neutron sensitivity \cd TEDs commonly used in combination with the Pulse-Height Weighting Technique (PHWT) are
replaced by high resolution position and energy sensitive radiation detectors consisting of two detection stages. Operated in time-coincidence, they allow one to apply the Compton principle in order to obtain also information on the incoming radiation direction~\cite{Everett77}. 
Thus, it becomes possible to disentangle whether the registered radiation stems from the capture-sample under study, which may indicate a true capture event, or if it comes from the surrounding environment, which would rather reflect a background event. The high energy resolution required for the proposed i-TED detection system also implies a superior performance for treating other background sources, such as intrinsic sample radioactivity, when compared to conventional BaF$_2$ TAC or C$_6$D$_6$ detectors.
One limitation of the proposed i-TED system is the attainable efficiency, which is about a factor of 3-4 lower than with existing C$_6$D$_6$ TED-detectors. This can be compensated to some extent by arranging four Compton modules in a compact geometry around the sample under study.  

A conceptual description of i-TED is given in Sec.~\ref{sec:design}. The performance of the proposed system for ($n,\gamma$) measurements under severe background conditions is investigated here by means of Monte Carlo simulations. Sec.~\ref{sec:performance} describes the capability of i-TED for disentangling spatially localized background sources. The intrinsic neutron sensitivity of i-TED is reported in Sec.~\ref{sec:ns}. The low detection efficiency of the proposed detection system implies that it has to be used in conjunction with the PHWT in order to make the capture detection probability independent of the particular decay path. The applicability of the PHWT to the proposed i-TED concept is reported in Sec.~\ref{sec:phwt}. A summary and outlook of the present work is presented in Sec.\ref{sec:summary}.

\section{i-TED conceptual design}\label{sec:design}
For the design of a detection system suitable for \ngr cross section measurements one has to take into account the neutron induced $\gamma$-ray background characteristic of this kind of experiment. 
Typically the overall neutron-induced $\gamma$-ray background has two components, which can be classified as $i)$\emph{ intrinsic neutron sensitivity} of the detection apparatus itself, which originates from neutrons scattered in the sample and captured prompt in the sensitive detection volume or in the detector structural materials, and $ii)$ the \emph{extrinsic neutron sensitivity} caused by stray neutrons that have been scattered in the sample and which are captured (normally after thermalization) in the surrounding structural materials like the concrete walls of the experimental hall. 
Each one of these background components has a different signature in the measured capture yield. The intrinsic neutron sensitivity of the apparatus is of concern mainly for the measurement of resonances where the neutron elastic scattering width is much larger than the capture channel. In those cases, a large intrinsic neutron sensitivity of the aparatus induces a strongly time of flight (TOF) dependent background which, if not properly corrected, can lead to an overestimated capture kernel or cross section for that particular resonance.
On the other hand, the neutron-thermalisation involved in the extrinsic neutron-sensitivity component leads to a background contribution, which is rather smooth in TOF. This has been nicely illustrated in a recent study~\cite{Zugec14}, where these two background components have been quantified by means of both systematic MC-simulations and dedicated measurements. The results indicate that using either C-fibre optimized or commercially (optimized) C$_6$D$_6$ detectors, the overall neutron-induced background over most of the primary neutron energy range (1eV-1MeV) is dominated by the extrinsic neutron sensitivity component of the facility (see Fig.6 and Fig.10 in Ref.~\cite{Zugec14}). Although the latter study~\cite{Zugec14} was made for the CERN n\_TOF facility, a comparable situation may be encountered at other TOF facilities~\cite{Ullmann05,Borella07}, where the instrumentation and experimental hall are also similar.

The method pursued in this work to provide a further level of background rejection is based on the implementation of radiation detectors with imaging capability, that enable the distinction between $\gamma$-rays arising from the capture-sample (true capture events) and $\gamma$-rays arising from elsewhere (background events).

There exist several ways to implement $\gamma$-ray imaging. One option could be to deploy a system based on one or several $\gamma$-ray cameras featuring a pinhole colimator~\cite{Anger58} or a coded-mask colimator~\cite{Zand92}. The main drawback of this approach is related to the very low detection efficiency and the use of massive collimators, which would induce a prohibitive intrinsic neutron sensitivity in ($n, \gamma$) measurements.
An alternative approach is to use a system based on Compton imaging~\cite{Everett77}; Electronic collimation allows one to enhance detection efficiency while reducing the amount of structural (dead) materials at the same time. 
Thus, the approach proposed in this work consists of Compton detection modules based on scintillation crystals arranged in a compact configuration and covering a large solid angle around the capture sample. Each module consists of a first detection layer (scatter detector), where the incoming radiation is expected to undergo just one Compton interaction. The remaining energy of the incident $\gamma$-ray is assumed to be fully deposited in a second thick detection layer (absorber detector). In order to be able to apply the Compton scattering law reliably, a good resolution both in energy and position becomes mandatory. This can be technically accomplished by using pixelated or monolithic fast scintillation-crystals with high scintillation photon yield, such as LaBr$_3$, LaCl$_3$~\cite{Moses05} or CeBr$_3$~\cite{Shah05}, coupled to thin photosensors such as arrays of avalanche photodiodes (see e.g.\cite{Conde14,Gonzalez13,Llosa13,McClish09}). The latter are also known as silicon photomultipliers (SiPMs) or multi-pixel photon-counters (MPPCs). Position-sensitive photomultiplier tubes (PS-PMTs) can be used for the absorber detectors~\cite{Pani09,Domingo09}, but in order to minimize the intrinsic neutron sensitivity the much lighter and thinner SiPM photosensors are preferred.

\begin{figure}[htbp!]
\flushleft
\centering
\includegraphics[width=1\columnwidth]{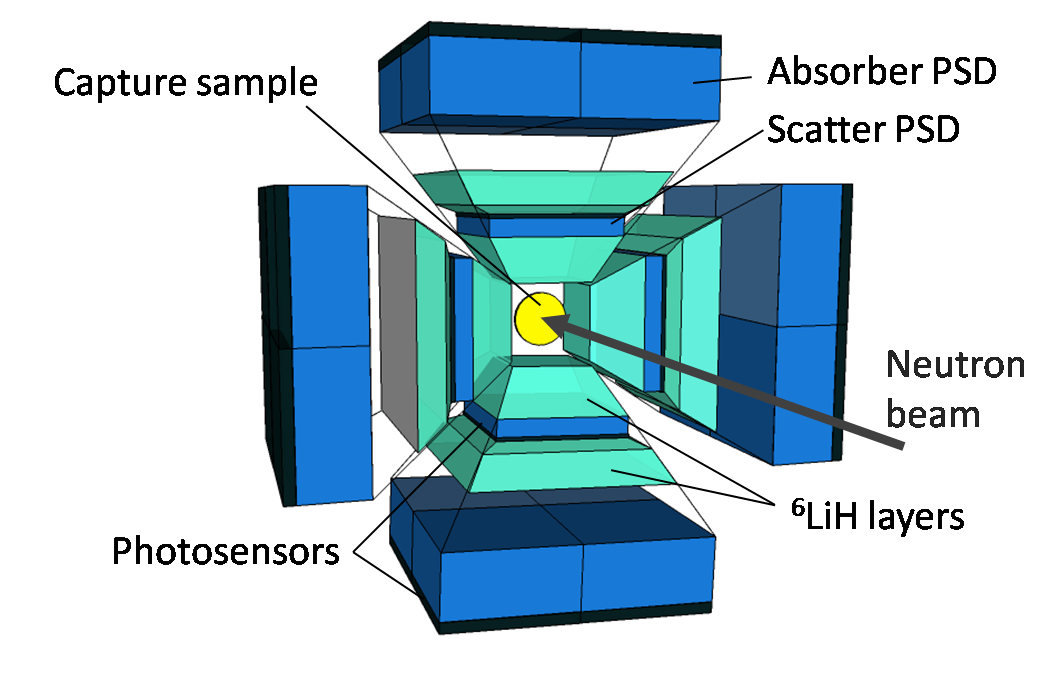}
\caption{\label{fig:ited} The cylindrical capture sample of 2~cm diameter in the center is surrounded by four thin scatter detectors and four thick absorber detectors configuring a compact Compton total energy detector with imaging capabilities (i-TED). The neutron beam impinges perpendicular to the sample surface.}
\end{figure}

In order to illustrate the working principle and performance let us consider an i-TED system consisting of four (scatter)
\lacl crystals with a thickness of 4~mm and a square area of 5$\times$5~cm$^2$ surrounded by four (absorber) crystals 12~mm thick, each one with an area of 10$\times$10~cm$^2$. Each one of the latter position sensitive absorber detectors can be built using four smaller detectors of 5$\times$5~cm$^2$. In this example, the distance between scatter and absorber detectors is of 1.7~cm. This distance defines the $\gamma$-ray efficiency of the system, but also its angular resolution and therefore its performance in terms of $\gamma$-ray imaging. Therefore, the optimal distance between scatter and absorber detectors needs to be adjusted for each particular detection set-up and it can be optimized according to the needs of each particular ($n,\gamma$) measurement, such as overall background level and magnitude of the cross-section of interest.
In order to reduce the intrinsic neutron sensitivity of the overall apparatus the scatter detectors are covered with $^6$LiH-layers (see Sec.~\ref{sec:ns}). This approach is similar to the one commonly employed to reduce the intrinsic neutron sensitivity of TAC-systems~\cite{Reifarth05,Guerrero09}. The proposed i-TED setup is schematically shown in Fig.~\ref{fig:ited}.

The applicability of this apparatus for the measurement of \ngr cross sections can be evaluated in terms of the extrinsic- and intrinsic-neutron sensitivity, which are reported in the two following sections.

\section{i-TED performance in terms of extrinsic neutron sensitivity}\label{sec:performance}
MC simulations are carried out using a simplified model for the background induced by neutrons scattered in the sample and captured in the surrounding environment. These calculations are described in the following. 

A cylindrical sample of gold, with a thickness of 1~mm and a diameter of 2~cm is used for
illustration purposes because gold is rather well known and commonly used in many time-of-flight (TOF) \ngr experiments as reference~\cite{Massimi10,Lederer12,Massimi14}.
In the simulation neutrons with a flat energy distribution in dE/E, i.e. iso-lethargic flux, from thermal up to 1~MeV impinge on
the center of the sample. Neutron transport and all possible neutron interactions are
included (capture, thermal, elastic and inelastic scattering and fission) using the
\textsc{Geant4}~\cite{Geant4} version (4.10.0.3) and the High-Precision neutron interaction libraries. Electromagnetic
processes are included by means of the Low Energy package.

In order to benchmark the performance with respect to existing systems both the i-TED system (see Fig.~\ref{fig:ited}) and two \cd detectors with a volume of 1~L are simulated. 

\begin{figure}
\flushleft
\centering
\includegraphics[width=\columnwidth]{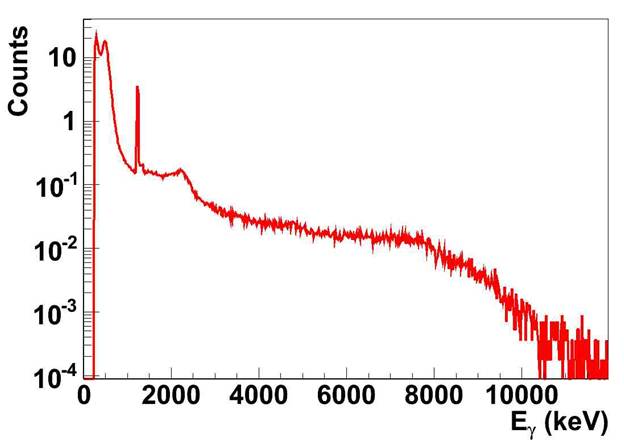}
\caption{\label{fig:e_bkg} $\gamma$-ray energy distribution implemented in the simulation for background events.}
\end{figure}

The extrinsic background depends on the characteristics of each particular measurement, including the capture sample itself, the surrounding materials and the experimental hall. In order to investigate the proposed method from a general perspective the background contribution from the extrinsic neutron sensitivity is modeled in the simulation as a source of $\gamma$-rays, which are randomly generated all over the surface of a 1~m radius sphere centered around the sample under study. 
These background $\gamma$-rays are generated with isotropic angular distribution and arbitrary multiplicity of 100 gamma-quanta per each neutron impinging on the sample. The energy distribution and intensity of the background varies from one facility to another, and for each particular sample under study. In the present study, just for illustration purposes, the energy distribution of the gamma-background included in the simulation corresponds to a measurement made at CERN n\_TOF~\cite{Guerrero13} with one BaF$_2$ detector, and is shown in Fig.~\ref{fig:e_bkg}.

\begin{figure}[htbp!]
\flushleft
\centering
\includegraphics[width=\columnwidth]{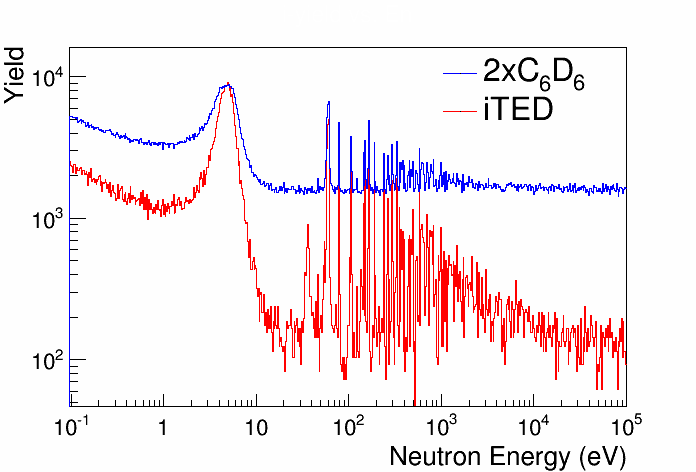}
\caption{\label{fig:auyield} Simulated capture yield for a 1~mm thick gold sample normalized at the 4.9~eV resonance. See text for details.}
\end{figure}

Thus, the two main $\gamma$-ray sources accounted for in the simulation are $\gamma$-rays originated by neutron capture in the gold sample and background radiation emitted around the detection system according to the aforementioned background generation model.
The Monte Carlo simulation of neutrons is a CPU very demanding task. For this study 1$\times$10$^7$ neutrons impinging on the gold sample are simulated for each detection setup. The obtained Au($n,\gamma$) capture yield as a function of the incident neutron energy is shown in Fig.~\ref{fig:auyield} for the two simulated set-ups, C$_6$D$_6$ and i-TED. Because the aim is to show the performance in terms of peak-to-background (signal-to-background) ratio, both spectra are normalized to the top of the 4.9~eV resonance. However, it is worth mentioning that the detection efficiency of the i-TED system is about a factor of four lower than that of the \cd detectors. A detailed comparison in terms of detection efficiency is provided in Section~\ref{sec:efficiency}.
The blue curve in Fig.~\ref{fig:auyield} shows the capture yield for the two \cd detectors including only an electronic threshold of 150~keV. The background level is largest in this case, which ultimately prevents the detection of weak resonances such as the one at 46~eV in $^{197}$Au+n.  The proposed i-TED detection system, also with a threshold of 150~keV, shows a superior performance in signal-to-background ratio when a cut in angle is applied to select $\gamma$-rays coming from the sample direction (red curve). The analysis methodology implemented for the i-TED simulated capture data is described below in Sec.~\ref{sec:imaging}. Using the ratio of the yield at 4.9~eV (peak) and at 20~eV (valley), an improvement of a factor $\sim$10 is obtained using i-TED with respect to the \cd detection system.

\section{i-TED performance in terms of intrinsic neutron sensitivity}~\label{sec:ns}
\begin{figure}[htbp!]
\flushleft
\centering
\includegraphics[width=\columnwidth]{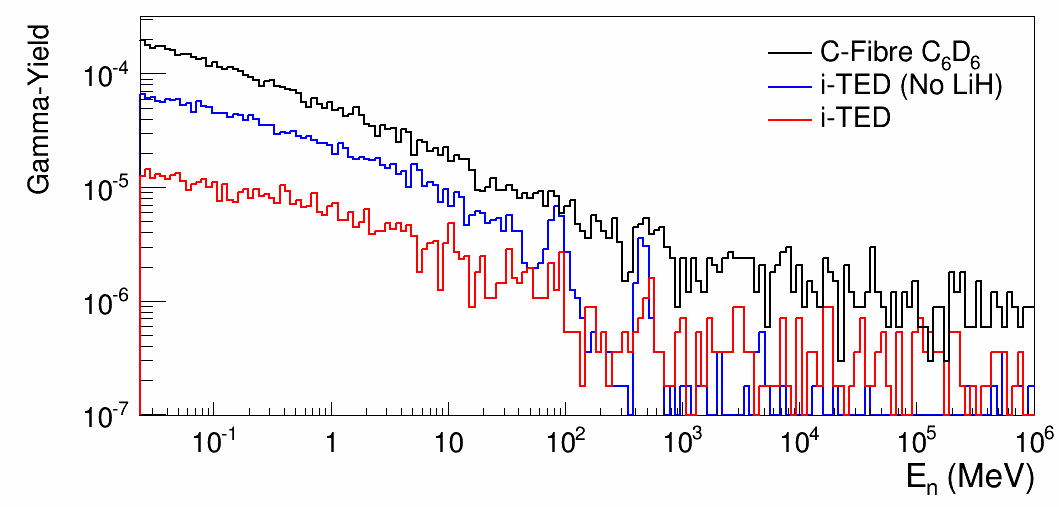}
\caption{\label{fig:cyield} Simulated detection gamma-yield for 1~cm thick C-sample using a set-up of two C-fibre optimized \cd detectors (black), i-TED without LiH-layers (blue) and i-TED with LiH-layers (red).}
\end{figure}

\cd is the most common liquid scintillation material employed for \ngr measurements using the PHWT due to the very low neutron capture cross section of carbon and deuterium, which directly lead to a very low intrinsic neutron sensitivity. Thus, when replacing the sensitive detection volume by another material, especially with an inorganic crystal, the first concern is obviously in terms of intrinsic neutron sensitivity. The neutron capture and scattering cross sections of both lanthanum and chlorine are much larger than those of carbon and deuterium. However, using the proposed method, the probability that a contaminant neutron capture event in the detector mimics a good Compton event compatible with a $\gamma$-ray coming from the sample position is very low. In order to demonstrate this, a simulation has been carried out using a carbon sample with a thickness of 1~cm. An optimised version of the C-fibre C$_6$D$_6$ detectors~\cite{Plag03} was included in the simulation in order to account for the effect of dead materials in state-of-the-art detectors. No surrounding background events (e.g. neutron capture in the concrete walls of the experimental hall) have been included in order to illustrate better the impact of the intrinsic neutron sensitivity itself. Given the small neutron sensitivity of both C-fibre C$_6$D$_6$ and i-TED it was necessary to generate 1$\times$10$^9$ neutron events in order to achieve sufficient counting statistics over the full neutron energy range.
The simulated spectra of detected $\gamma$-rays as a function of the neutron energy for both \cd and i-TED detectors are shown in Fig.~\ref{fig:cyield}. The rather smooth trend of these spectra reflects mainly neutron scattering events in the carbon sample, which are subsequently captured in the detection volume (\cd or \lacl) or in the detector structural detector materials.
 
The $^6$LiH layers have a twofold impact in reducing the intrinsic neutron sensitivity of i-TED. On one hand, they produce a moderation and absorption of the neutrons scattered in the C-sample, in a similar fashion as it happens in measurements with TACs~\cite{Reifarth05,Guerrero09}. On the other hand, the $^6$LiH layers produce a defocussing effect in the spatial distribution of the neutron source (originally confined to the sample volume), which in combination with the aforedescribed imaging algorithm helps to further suppress the contribution of contaminant neutron capture events in the detection set-up.
  
In summary, it can be concluded that the background induced in i-TED due to scattered neutrons which are captured in the detector itself is a factor of five to ten lower than for state-of-the-art C$_6$D$_6$ detectors in the energy range between 1~eV and several hundreds of keV.

\section{Compton imaging analysis method}\label{sec:imaging}
At variance with other Compton camera applications~\cite{Azevedo13,Kataoka13} the prompt $\gamma$-ray cascades following neutron capture events show a rather broad energy spectrum. One possibility for background rejection would be to implement a tomographic algorithm (see e.g. Ref.~\cite{Barrett09}), which takes into account the different $\gamma$-ray spectra between neutron capture events in the sample and neutron-induced $\gamma$-ray backgrounds from the surroundings. Also very promising energy-imaging spectral deconvolution algorithms have been developed recently~\cite{Xu07,Wang12}. However, in this first demonstration stage, a more simple approach is followed using an analytic back-projection method~\cite{Wilderman98} in combination with the fact, that the geometry of the set-up (sample position, size and distance to the i-TED detector) is known by construction. Thus, the compatibility of the measured energy and positions in scatter and absorber detectors is checked on an event-by-event basis for the sample position using the Compton formula and the aforementioned geometry constraints. 
The equation describing the intersection of the Compton cone with the sample is given by

\begin{equation}\label{eq:intersection}
\begin{aligned}
\big( n_x (x_s - a_x) + n_y (y_s - a_y) + n_z (z_s - a_z) \big)^2 = \\
\shoveleft{\cos^2\theta \big((x_s-a_x)^2 + (y_s - a_y)^2 + (z_s - a_z)^2\big),}
\end{aligned}
\end{equation}

where $n_s$ are the components of a unit vector along the cone axis (the vector between the first and second interactions), $a_s$ are the coordinates of the first interaction in the scatter detector, ($x_s, y_s, z_s$) is the position of the sample (true capture $\gamma$-ray source) and $\theta$ is the Compton scattering angle. The latter can be determined from the energy deposited in scatter and absorber detectors by applying the Compton formula,

\begin{equation}\label{eq:compton}
\cos\theta = 1 + \frac{511}{E_g} - \frac{511}{E_2},
\end{equation}

where $E_g$ is the energy of the incident $\gamma$-ray, which is assumed to correspond to the sum of the energy in scatter $E_1$ and absorber detector $E_2$.

\begin{figure}[htbp!]
\flushleft
\centering
\includegraphics[width=\columnwidth]{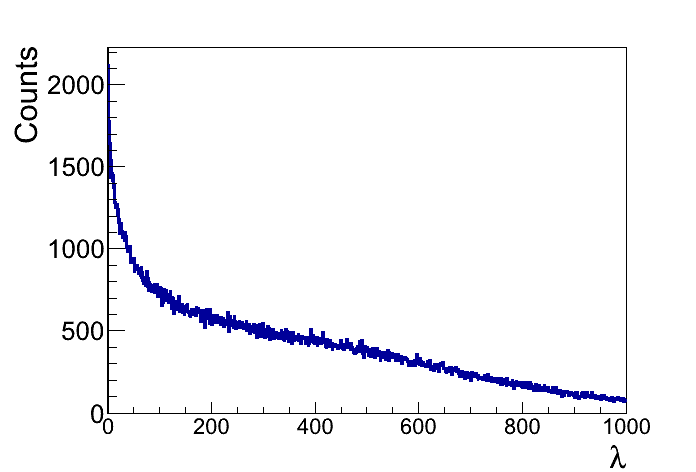}
\caption{\label{fig:lambda} $\lambda$-distribution for applying a cut in space-domain.}
\end{figure}

Thus, in order to check the compatibility of the measured radiation with the sample position, the quantity
\begin{equation}\label{eq:delta}
\begin{aligned}
\lambda = \big( n_x a_x + n_y a_y + n_z a_z \big)^2 - \\
\shoveleft{(1 + 511/(E_1+E_2) - 511/E_2)^2 \big(a_x^2 + a_y^2 + a_z^2\big),}
\end{aligned}
\end{equation}

can be employed. The sample position defines the origin of the coordinates system, i.e. ($x_s,y_s,z_s) = (0,0,0)$. The $\lambda$ distribution obtained for the MC simulation described in the previous section is shown in Fig.~\ref{fig:lambda}. Low $\lambda$ values correspond to detected $\gamma$-rays which fulfill the intersection condition (eq.~\ref{eq:intersection}) and, therefore, have a high probability to arise from the capture sample (true capture events). On the other hand, large $\lambda$ values correspond to events in which the Compton cone does not overlap with the sample, and thus rather indicate a background event probably induced from contaminant neutron captures in the surrounding materials or walls.
Applying a selection or cut to choose low $\lambda$ values one can effectively reject many $\gamma$-rays which impinge on the system from the surrounding (background events). A cut in $\lambda \leq 90$ gives the red yield-curve (labeled as i-TED) in Fig.~\ref{fig:auyield}. A more stringent background rejection could be achieved by using a lower $\lambda$-cut at the cost of statistics.

\section{Applicability of the PHWT to i-TED}\label{sec:phwt}
Because the $\gamma$-ray detection efficiency is small, the PHWT~\cite{Macklin67} needs to be applied to the measured response function in order to make the capture-cascade detection efficiency independent of the actual decay path. The PHWT is based on the two assumptions of $i)$ low $\gamma$-ray detection efficiency, and $ii)$ $\gamma$-ray detection probability proportional to the $\gamma$-ray energy. Provided that these two conditions are well fulfilled, the probability to register a neutron capture event, $\varepsilon_c$, becomes proportional to the total capture-cascade energy $E_c$, which is a constant value, and thus it does not depend any more on the particular $\gamma$-ray detected,

\begin{equation}\label{eq:Ec}
\varepsilon_c = 1 - \prod^{m}_{j=1} (1 - \varepsilon_{\gamma,j}) \stackrel{i)}{\simeq} \sum^{m}_{j=1} \varepsilon_{\gamma,j} \stackrel{ii)}{\simeq} \alpha E_c.
\end{equation}

Low efficiency is mandatory to ensure that at most one $\gamma$-ray is registered from every capture cascade with arbitrary multiplicity $m$ . This condition allows for the first approximation $i)$, which is generally well fulfilled using low efficiency C$_6$D$_6$ detectors, an statement which also applies for the proposed i-TED system (see Sec.~\ref{sec:efficiency}). 
The proportionality condition $ii)$ is, in general, not directly fulfilled by any existing radiation detector. However, it can be achieved to a high-level of accuracy by means of a software modification of the detector response function R(E). As reported in e.g.\cite{Abbondanno04,Borella07}, one can determine an energy-dependent weighting-function $W(E)$, such that the weighted sum of the response function for a $\gamma$-ray $j$, $R_{i,j}$, becomes proportional to its energy $E_{\gamma,j}$,

\begin{equation}\label{eq:wsum}
\sum_i W_i R_{i,j} = \alpha E_{\gamma,j}.
\end{equation}

The common approach to determine the weighting function is based on Monte Carlo simulations of the detection system response for a series of $\gamma$-ray energies in the energy-range of interest, i.e. from few keV up to the neutron capture energy $E_c$ of the isotope of interest. Typically the weighting-function is well approximated by a 4- or 5-degree polynomial, $W_i = \sum_{k=0}^{4,5}a_k E_i^k$. The values of the polynomial coefficients $a_k$ can be derived from a least-squares minimization,
\begin{equation}\label{eq:min}
min \sum_j \left( \sum_i a_k E_i^k R_{i,j} - E_{\gamma j} \right)^2.
\end{equation}

\begin{figure}[htbp!]
\flushleft
\centering
\includegraphics[width=0.7\columnwidth]{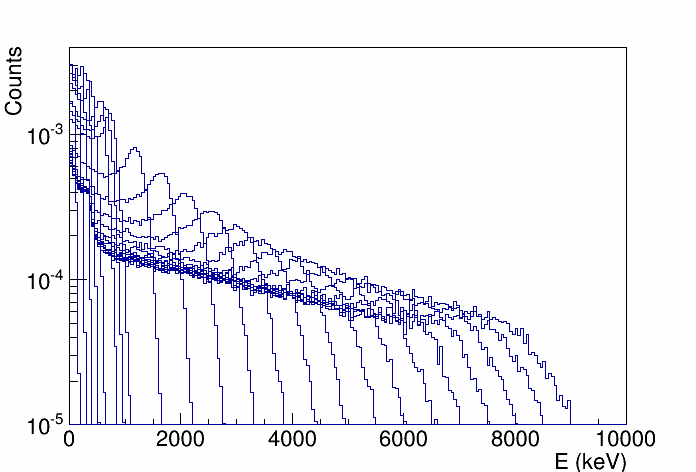}
\hfill
\includegraphics[width=0.7\columnwidth]{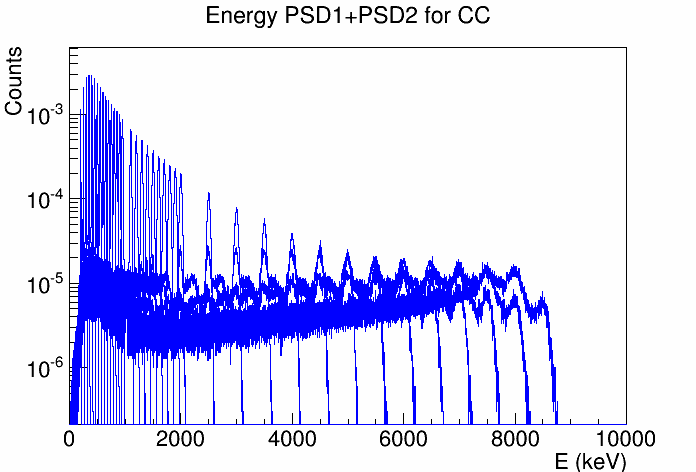}
\caption{\label{fig:rfs} Response functions (5~keV/bin) for 25 $\gamma$-ray energies from 100~keV to 8.5~MeV for the two C$_6$D$_6$ detectors (top) and for the i-TED (1~keV/bin) system (bottom).}
\end{figure}

For the present study 25 $\gamma$-ray energies in the range from 100~keV up to 8.5~MeV were simulated for both \cd and i-TED detection systems. These response functions have been convoluted by the typical instrumental resolution at 662~keV of 20\% for the liquid C$_6$D$_6$ detectors and 4\% for the inorganic LaBr$_3$ scintillators. The simulated and convoluted response functions are displayed in Fig.~\ref{fig:rfs}. Since the weighting function calculated for i-TED is going to be applied to the raw data processed with the back-projection cut described in Sec.~\ref{sec:imaging}, the response functions shown in Fig.~\ref{fig:rfs}-Right include also the $\lambda \leq 90$ selection.

\begin{figure}[htbp!]
\flushleft
\centering
\includegraphics[width=0.7\columnwidth]{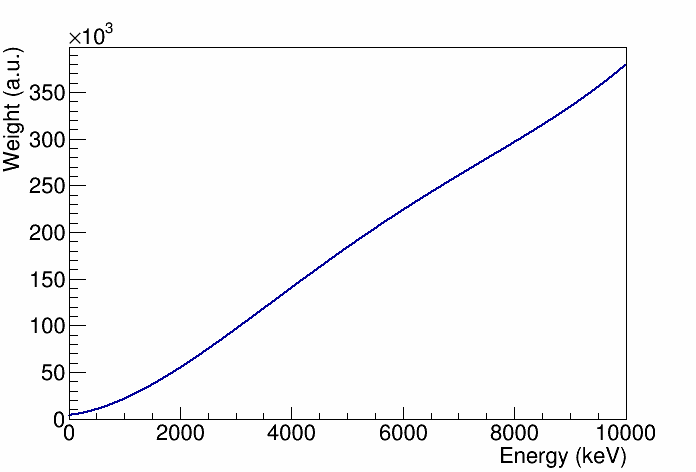}
\hfill
\includegraphics[width=0.7\columnwidth]{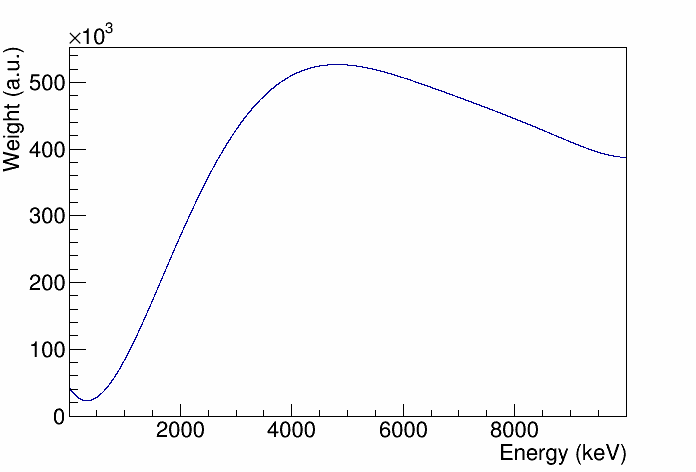}
\caption{\label{fig:wfs} Weighting functions obtained for the C$_6$D$_6$ detection system (top) and for i-TED (bottom).}
\end{figure}

After implementing a software algorithm which performs the minimization described by eq.~(\ref{eq:min}) one obtains the weighting functions shown in Fig.~\ref{fig:wfs} for both detection systems.

\begin{figure}[htbp!]
\flushleft
\centering
\includegraphics[width=0.7\columnwidth]{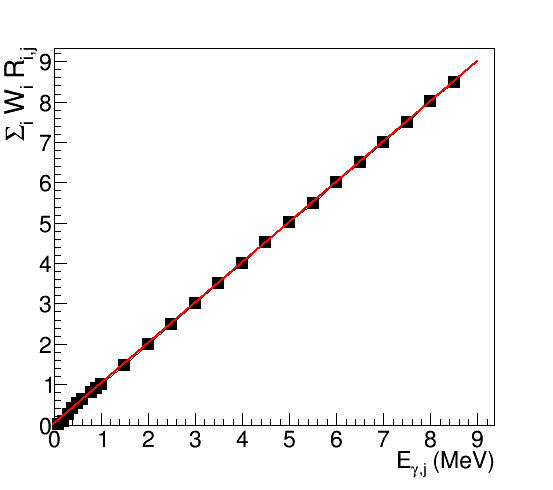}
\hfill
\includegraphics[width=0.7\columnwidth]{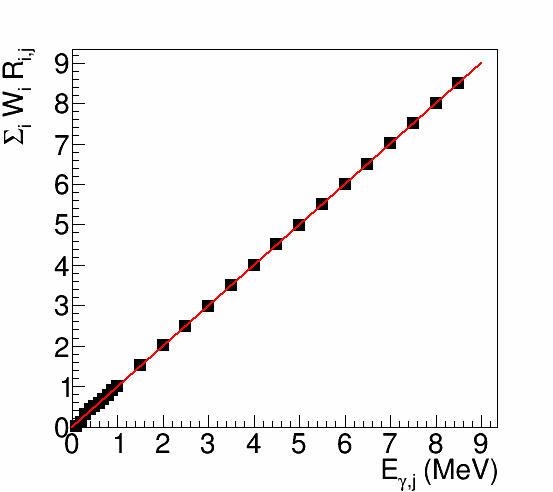}
\caption{\label{fig:wsum} Integral of the weighted response as a function of the simulated $\gamma$-ray energy for the C$_6$D$_6$ detectors (top) and for the i-TED system (bottom). The solid red line shows the behavior of an ideal total-energy detector with a perfect proportionality between efficiency and energy.}
\end{figure}

In the case of the simpler response function of the C$_6$D$_6$ detectors, characterized by the Compton continuum, a 5-degree polynomial was sufficient to fulfill the proportionality condition (eq.\ref{eq:wsum}). On the other hand, a 7-degree polynomial is needed for the i-TED weighting function in order to fulfill the proportionality condition with a similar level of precision. This is expected also from the contribution of both Compton and full-energy events to the response function of inorganic scintillation detectors.
After applying the calculated WFs (Fig.~\ref{fig:wfs}) to the simulated responses (Fig.~\ref{fig:rfs}) both weighting functions perform similarly well. This is demonstrated in Fig.~\ref{fig:wsum}, which shows the weighted response function as a function of the $\gamma$-ray energy.

\begin{figure}[htbp!]
\flushleft
\centering
\includegraphics[width=\columnwidth]{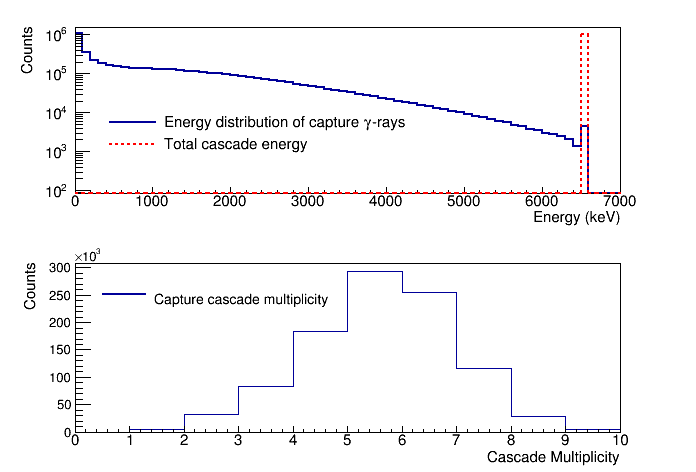}
\caption{\label{fig:casc} (Top panel) The spectrum (solid blue line) shows the energy distribution of the $\gamma$-rays emitted after neutron capture in gold. The dashed line at 6.512 MeV shows the total cascade energy recorded for the $1\times10^6$ events simulated. (Bottom panel) Multiplicity distribution of the simulated gamma-quanta. }
\end{figure}

In order to estimate the systematic uncertainty contribution of the calculated weighting functions in a real neutron capture experiment, a computational approach similar to the one described in Ref.\cite{Abbondanno04} has been applied. To this aim, $1\times10^6$ neutron capture events in gold are simulated with isotropic emission of all the gamma-quanta in each cascade. Unfortunately, the \textsc{Geant4} code (version 4.10) used for these calculations does not conserve the total energy in the generation of radiative capture cascades. Therefore, a simplified external event generator~\cite{Mendoza14} had to be used for this purpose. The energy and multiplicity distribution of the simulated capture cascades in the gold sample are shown in Fig.~\ref{fig:casc}.

\begin{figure}[htbp!]
\flushleft
\centering
\includegraphics[width=0.49\columnwidth]{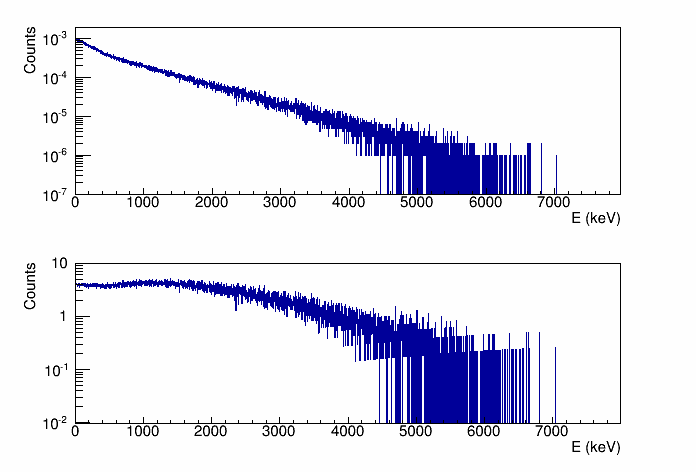}
\hfill
\includegraphics[width=0.49\columnwidth]{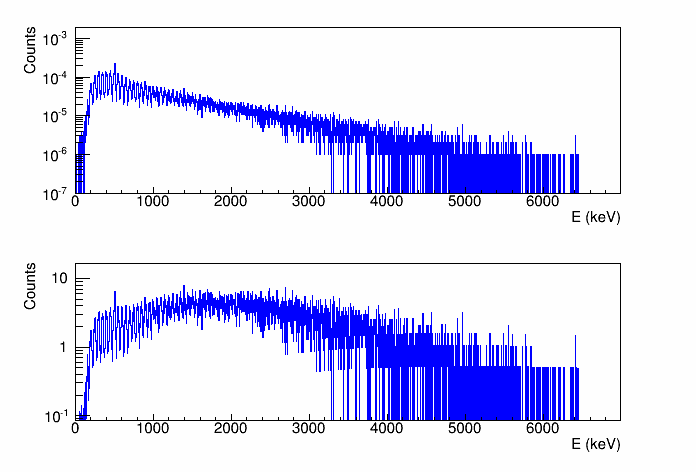}
\caption{\label{fig:wrf_au} Raw (top) and weighted (bottom) response functions for neutron capture events in gold using the C$_6$D$_6$ detection system (left) and i-TED (right).}
\end{figure}

\begin{table}[!htbp]
\caption{Weighted sum of the response functions for neutron capture events in gold (2$^{nd}$ column) and weighted sum normalized to the number of simulated capture events times the capture energy E$_c$ = 6.512 MeV (3$^{rd}$ column). The last 4$^{th}$ column shows the systematic deviation with respect to the ideal case.}\label{tab:wrf}
\begin{center}
\begin{tabular}{cccc}
\hline
Detection &  &  & Sys.\\
system & $\sum_i W_i R_{i,c}$ & $\sum_i W_i R_{i,c}/(N_{ev} E_c)$ & deviation\\
\hline
2$\times$ C$_6$D$_6$ & 6.389(6)$\times 10^{9}$ & 0.9812(9) & 1.9\%\\
i-TED & 6.36(2)$\times 10^{9}$ &  0.977(3) & 2.3\%\\
\hline
\end{tabular}
\end{center}
\end{table}

The raw $R_{i,c}$ and weighted $W_i R_{i,c}$ response functions obtained for the simulation of 1$\times$10$^{6}$ capture events in gold are shown in Fig.~\ref{fig:wrf_au}. The relevant information, however, is given in table~\ref{tab:wrf} which reports the integrated value of the weighed response function for both detection systems. Deviations of around 2\% are found in both cases, which demonstrate that the PHWT can be applied to both types of detection systems with similar level of accuracy.

\subsection{Weighted capture yields}\label{sec:wau}
\begin{figure}[htbp!]
\flushleft
\centering
\includegraphics[width=\columnwidth]{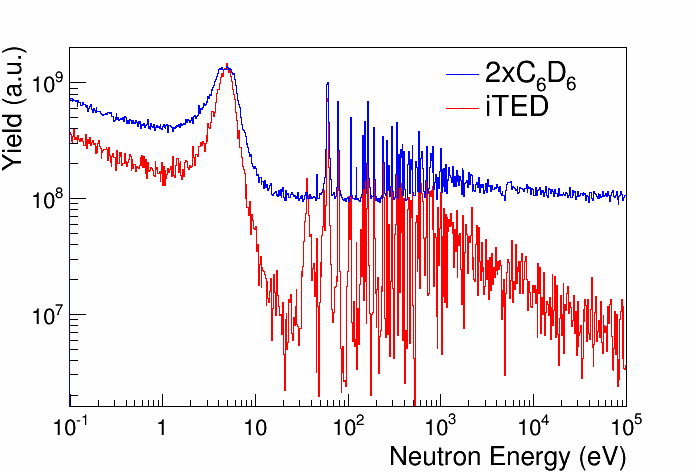}
\caption{\label{fig:wauyield} Simulated capture yield as a function of the neutron energy for both detection systems C$_6$D$_6$ and i-TED. The weighting functions applied are those reported in Sec.~\ref{sec:phwt}.}
\end{figure}

As discussed above, in order to determine the neutron capture cross section the raw capture yield has to be weighted with the corresponding weighting function to make it independent of the particular cascade path or registered energies. After implementing the weighting functions calculated in the preceding section to the raw-capture yield simulated for the gold sample (see Fig.~\ref{fig:auyield}) one obtains the weighted capture yield shown in Fig.~\ref{fig:wauyield}.
Applying the same figure of merit as before, i.e., comparing the capture yield at the top of the 4.9~eV resonance (peak) to the value at 20~eV (valley) one obtains an improvement in peak-to-background ratio, which ranges between a factor of 12 to 15 in favor of i-TED with respect to the C$_6$D$_6$ detectors.

\subsection{Efficiency}\label{sec:efficiency}

\begin{figure}[htbp!]
\flushleft
\centering
\includegraphics[width=\columnwidth]{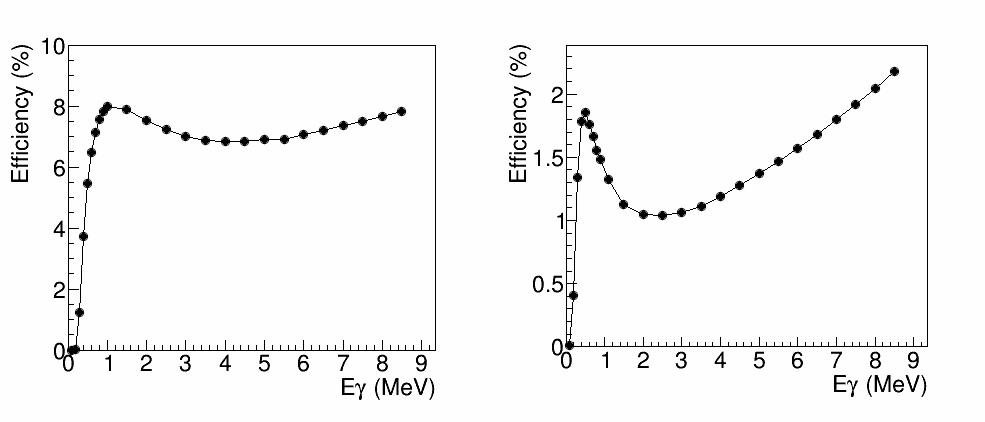}
\caption{\label{fig:eff} $\gamma$-ray detection efficiency as a function of the $\gamma$-ray energy for the system of two C$_6$D$_6$ detectors (left) and i-TED (right).}
\end{figure}

The $\gamma$-ray detection efficiency for i-TED is presented in Fig.~\ref{fig:eff} as a function of the $\gamma$-ray energy. This efficiency includes a low-energy detection threshold of 150~keV. For the sake of comparison, the same diagram is also shown for the conventional C$_6$D$_6$ detectors. For the presented i-TED configuration, at low $\gamma$-ray energy the efficiency almost reaches 2\% at 500~keV. With the two C$_6$D$_6$ detectors the efficiency is of 8\% at 1~MeV, a factor of four more than with i-TED.
One possibility to enhance detection efficiency in i-TED is to reduce the distance between scatter- and absorber-detectors. This procedure, however, leads to a decrease of the angular resolution and therefore, to a worse performance in terms of background rejection. As discussed in Sec.~\ref{sec:design} the final i-TED configuration, and in particular the distance between scatter and absorber detectors, can be adjusted according to the needs of each particular measurement, such as the magnitude of the cross section to be measured, or the level of the background in the measurement.

\section{Summary and outlook}\label{sec:summary}
In view of the large neutron luminosities at present and future neutron time-of-flight facilities, and the correspondingly large neutron-induced backgrounds, parallel developments both in detection methods and instrumentation, capable to cope with the severe background conditions, are definitely needed.

This article presents a novel experimental approach to measure radiative neutron capture cross sections, which is aimed at enhanced detection sensitivity under severe background conditions. The proposed detection system is based on the use of fast inorganic scintillators arranged in a compact Compton camera configuration around the sample under study. Thus, it becomes possible to determine, on an event-by-event basis, a Compton cone of possible incoming directions for each registered $\gamma$-ray. Such angular information can be compared with the sample geometry and position, which are known by set-up construction. This allows one to reject a large portion of the incoming background $\gamma$-rays, mainly those arising from the surroundings of the detection system. Monte Carlo simulations presented in this work show the capability of the proposed technique for background suppression, and correspondingly a large improvement in peak-to-background ratio when compared to state-of-the-art C$_6$D$_6$ detectors. 

Future work in this research line concerns mainly two aspects. The first one is the replacement of the simplified background model assumed in this study by a full simulation of a neutron capture experiment, including also neutron propagation, transport and interactions not only in a reference capture sample, but also in the surrounding materials and walls. This requires a large computing infrastructure and long simulation time. Such a MC study will be helpful in order to further optimize the conceptual design of i-TED, namely the scatter and absorber detector thicknesses and distances, or a range of them. The second aspect concerns proof-of-principle measurements, which should be ideally carried out with a Compton $\gamma$-ray imager at a time-of-flight facility, in order to experimentally validate the proposed technique and to quantify its performance under real experimental conditions. First steps in this direction have been recently carried out using a pinhole based $\gamma$-ray camera~\cite{Perez15}, which already demonstrate the potential of $\gamma$-ray imaging for background suppression in ($n, \gamma$) measurements.

\section*{Acknowledgment}
The author acknowledges helpful discussions with C.~Guerrero, J.E.~Guillam and J.L.~Tain. This work was supported by Spanish \textit{Ministerio de Econom{\'{\i}}a y Competitividad} under Grant No. FPA2011-24553.

\bibliography{bibliography}

\end{document}